\documentstyle[aps,preprint,prl,floats]{revtex}
\tightenlines
\begin{document}
\title{Point defects, ferromagnetism and transport in calcium hexaboride}
\author{ R. Monnier$^{1}$ and B. Delley$^{2}$ }
\address{$^1$Laboratorium f\"ur Festk\"orperphysik, ETH-H\"onggerberg,
CH-8093 Z\"urich, Switzerland}
\address{$^2$Paul Scherrer Institut, CH-5232 Villigen PSI, Switzerland}
\maketitle
\begin{abstract}
The formation energy and local magnetic moment of a series of point
defects in CaB$_6 $ are computed using a supercell approach within
the generalized gradient approximation to density functional theory.
Based on these results, speculations are made as to the influence of
these defects on electrical transport. 
It is found that the substitution of Ca by La does not lead to the
formation of a local moment, while a neutral B$_6 $ vacancy carries a
moment of 2.4 Bohr magnetons, mostly distributed over the six nearest-neighbour 
B atoms. A plausible mechanism for the ferromagnetic ordering
of these moments is suggested. Since the same broken B-B bonds appear on the
preferred (100) cleavage planes of the CaB$_6$ structure, it is argued
that internal surfaces in polycrystals as well as external surfaces
in general will make a large contribution to the observed magnetization.
\end{abstract}


Recent experiments on alkaline earth hexaborides have
revealed an extreme sensitivity of their physical
properties to stoichiometry and impurity content.
A big surprise has been the observation of high temperature weak
ferromagnetism in CaB$_6 $, SrB$_6 $ and BaB$_6 $ lightly doped
with lanthanum and thorium, as well as in the isovalent
substitutional alloy Ca$_{0.995}$Ba$_{0.005}$B$_6$ \cite{young},
soon followed by
the discovery of the same phenomenon in nominally pure CaB$_6 $
\cite{vonl}, SrB$_6 $ \cite{naga},and BaB$_6 $ \cite{ottp}. The
authors of ref. \cite{vonl}  further found that when CaB$_6 $
is grown from a calcium-rich mixture of the elements, it becomes
very weakly \emph{paramagnetic}, while its low temperature (T$<$80 K)
resistivity is increased by more than two orders of magnitude.
The former observation suggests that
the ferromagnetism of the nominally pure systems is related
to the presence of vacancies on the metal sublattice and/or to the
resulting intrinsic (hole) doping.
In an attempt to quantify these apparent correlations,
Morikawa et al. \cite{morikawa} have performed a systematic study
of the magnetization and electrical resistivity of
CaB$_6$  samples synthesized from high purity CaO and B at
different temperatures between 1200 ${^o}$C and 1500 ${^o}$C maintained
between 1 and 24 h. Their results are rather
puzzling: samples grown under the same
conditions (1500 ${^o}$C, 6h) show magnetizations differing by a factor
of ten, and there is no connection whatsoever between the growth
conditions, the resistivity and the magnetization. Our investigations
suggest that the observed difference in magnetic moments
is due to the fact that their measurements were
performed on pressed powder pellets, with empty spaces of varying sizes
between the grain boundaries, in contrast to those of refs.
\cite{young,vonl,naga,ottp}, which were made on single crystals.

Electronic structure calculations in the local density (LDA) or
generalized gradient (GGA) approximation
yield a semimetallic ground state for all stoichiometric alkaline
earth hexaborides, with a small overlap between the valence and the
conduction band at the X point of the Brillouin zone
\cite{mass,pick,mass1}. Building on this feature, Zhitomirsky et al.
\cite{zhito} and later two other groups \cite{balents,gorkov}
have put forward a model according to which
the above ground state is unstable with respect to the formation
of a spin triplet exciton condensate and, upon doping, evolves into
a ferromagnetic state with a small spontaneous moment and a high Curie
temperature, in line with the experimental observations.
A more realistic calculation of the single-particle excitation 
spectrum of CaB$_6 $, based on the so-called
\emph{GW} approximation \cite{hedin}, suggests that this compound
is a conventional semiconductor, with a rather large band gap of 0.8 eV
\cite{kelly}. This prediction has very recently been confirmed 
by bulk-sensitive x-ray absorbtion and emission experiments 
at the boron K-edge \cite{denli}, which definitely rules out the model of refs. 
\cite{zhito,balents,gorkov}.

In this letter, we investigate the possibility that
the observed spontaneous magnetization
is localized at imperfections in the CaB$_6$ lattice and qualitatively
discuss the effect of these imperfections on the electrical
conductivity. Specifically, we perform selfconsistent electronic structure
calculations for a number of point defects centered in a
3x3x3 periodically continued supercell, which amounts to  a defect
concentration of 3.7 \%. For each of them we compute the formation
energy and the value of the local moment, two quantities for which
the GGA used here should give reasonable results, since they are ground
state properties \cite{technical}. Our results are summarized in
Table~\ref{param}. The formation energy is defined as the difference
between the binding energy per supercell of the defected crystal and
that of pure, stoichiometric CaB$_6 $. Geometrical relaxation effects are
small, except in the case of the point-symmetry-breaking single boron vacancy, 
where the distance to the nearest B-octahedron along the broken 
bond is reduced by 0.1~\AA.

We first consider the defects on the metal sublattice.
When a neutral Ca atom is removed from the crystal, it takes along
two valence electrons, and the vacancy left behind acts as an
acceptor. If the stoichiometric compound is a conventional
semiconductor \cite{kelly}, holes will be created near the top of
the valence band, which belong to orbitals localized on the individual
B$_6$ octahedra \cite{mass} and are therefore expected to contribute
little to the dc conductivity.

The single crystals studied in refs. \cite{young,vonl} were
grown from mixtures of the pure elements in a liquid aluminium flux,
and the question arises, whether Al atoms were incorporated during the
growth process. As seen in Table~\ref{param}, it is indeed energetically
favourable for an Al atom to fill a Ca vacancy when the crystal is
grown from the gas phase.
When the ``atom reservoir'' consists of liquid Al and Ca, as is the
case over most
of the temperature range of interest here (923 K - 1723 K \cite{ott}),
the difference between the heat of vaporization of
liquid aluminium  and that of liquid calcium at the temperature of the
reaction has to be added to the formation energy. A lower bound for these
two quantities is given by their values at the respective boiling
points of the two elements :$\sim$2.96 eV/atom for Al and
$\sim$1.67 eV/atom for Ca (at one atmosphere) \cite{rubber}.
For the same reason, the heat of vaporization of liquid Ca must be
subtracted from the calculated Ca-vacancy formation energy\cite{cao}.
Once these two corrections have been made, the formation energies for
the two defects become $\sim$5.0 eV.

At constant volume, the equilibrium defect concentration is given by
\begin{equation}
	c = \exp \left( - \frac{E_{\rm Form}+\Delta F_{\rm vib}}{k_{\rm B} T}\right), \label{conc}
\end{equation}
where $\Delta$F$_{\rm vib}$ is the difference between the vibrational
free energy of the crystal with a defect and that of the perfect one.
A microscopic calculation of this difference is prohibitive, and the
standard treatment \cite{dekker}, based on a description of the solid as
a collection of independent Einstein oscillators, predicts  that, in the
limit of high temperatures, it leads to a T-independent factor
in  the concentration. For vacancies, this factor is
typically of order ten. For a light impurity (Al) substituting
a heavier host atom (Ca), it will, in general, be smaller than one.
At T=1723 K and from the formation energies alone,
we find, for crystals grown in an aluminium flux, equilibrium
concentrations of the order of $\sim$10$^{-14}$. This is 10 orders of magnitude
less than the density of \emph{negative} charge carriers
deduced from transport measurements\cite{gianno}, and we have to conclude 
that the negative carriers are
either due to impurities in the starting materials or to defects
in the boron network (see below), if the growth took place under
conditions of thermal
equilibrium. For the compound synthesized by borothermal reduction of
CaO, the same approximation leads to an equally low vacancy concentration of
$\sim$10$^{-13}$\cite{cao}.

As seen in Table~\ref{param}, the substitution of Ca by La increases the
binding energy of the compound\cite{gain}.
This remains the case even
after correcting by the difference in heats of vaporization of La and
Ca ($\sim$1.81 eV/atom), and all the lanthanum in the flux should
therefore be incorporated. According to our calculation, there is no
magnetic moment associated with the La impurity. This is in contrast
to the result of Jarlborg \cite{jarl}, who finds a
moment of the order of 0.1 $\mu_{\rm B}$ in a similar study for La in
SrB$_6 $ within the LDA. Our method applied to that system again yields a 
vanishing moment and we have no explanation for the origin of the difference 
between the two calculations.

The last impurity we  have considered on the metal sublattice is boron,
whose presence could explain the metal-deficiency reported
by many experimental groups. According to Table 1, the corresponding
formation energy is  $\sim$1 eV higher than for the  substitution of Ca
by Al, which, combined with the higher cohesive energy of solid boron
makes it very unlikely for this defect to form under conditions
of thermal equilibrium.

Because of the strong covalent B-B bonds, the vacancy formation energy on
the boron sites is expected to be much higher than on the metal sites.
This is indeed what we find in our calculations. As seen in Table~\ref{param}
the difference between these two energies is considerably reduced when proper
account is taken of the growth procedure: 
under the conditions used in refs. \cite{young,vonl,morikawa}, every B atom 
lost by the compound condenses into solid boron. The latter has a cohesive energy 
of $\sim$5.9 eV/atom \cite{bausch}.
From this, its binding energy at 1723 K is obtained by subtracting
the change in vibrational free energy over the corresponding 
temperature interval,
which we approximate by its upper bound 3\emph{k$_{\rm B}$T}=0.47 eV/atom.
So, at least $\sim$5.4 eV are gained for
every single boron vacancy formed, and therefore the net energy cost
per vacancy drops from  $\sim$11.0 eV for the compound grown from
the gas phase, to $\sim$5.6 eV, leading to an equilibrium vacancy
(donor) concentration  of the order of 10$^{-15}$/unit cell at the quoted
temperature. This is 11 orders of magnitude lower the measured 
carrier concentration \cite{gianno}, and it is unlikely that the
change in vibrational free energy upon removing a B atom from the crystal
can account for such a large difference.

The loss of complete boron octahedra by the structure has
been invoked by Noack and Verhoeven \cite{noack} in order to explain their
gravimetric data on zone refined LaB$_6$. According to our 
calculations, this process is energetically more favourable than the 
creation of an equivalent number of well separated, single boron vacancies,
but the computed formation energy of 18.3 eV implies that, unless 
the resulting 
void is stabilized by another factor, like eg a large impurity cation 
or a Ca vacancy acting as an acceptor and taking away electrons from
interoctahedra bonding orbitals, it will not be formed under
thermal equilibrium conditions. What makes it interesting, however,
is the large magnetization it carries (see Table~\ref{param}).The moment is
mainly localized on the six neighbouring boron octahedra, most of it
(6 x 0.24 $\mu_{\rm B}$) in the dangling bonds from
the atoms immediatly adjacent to the void; the next shell of boron
atoms holds 24 x0.015 $\mu_{\rm B}$, while the 6 B atoms furthest away from
the defect carry 0.04 $\mu_{\rm B}$ each; finally, every adjacent Ca has a
moment of 0.04 $\mu_{\rm B}$.
A void concentration of the order of 10$^{-4}$ would suffice to
account for the largest saturation magnetization observed in
ref.\cite{young}. A possible explanation for the \emph{ordering} of these
moments can be obtained if one makes the reasonable assumption that,
in the presence of compensating cation vacancies and/or impurities,
a B$_6$ vacancy can not only be neutral but also positively
charged. In this case a  ``double exchange'' \cite{pwa,degennes} 
can take place between magnetic clusters formed by a ``void'' and its 
nearest neighbour boron octahedra and Ca ions,
provided they are in different charge states. The low
concentration of these clusters would require them to form a  loosely connected
network confined to a limited region of the crystal, for ferromagnetism to occur also
at the lowest temperatures.The existence of an
inhomogeneous ferromagnetic state in La doped CaB$_6$ has already been
proposed by Terashima et al.\cite{tera}.

A neutral B$_6$ void is the crudest approximation to the crystal
surface, which is known to cleave in the [100] plane through the
breaking of interoctahedra B-B bonds. Our results immediatly lead to
the conclusion that the outermost layer of CaB$_6$ will be magnetically
ordered, with a sizeable moment per unit cell. This suggests that the
very large magnetization observed in polycrystalline samples\cite{morikawa}
is located at the surface of the crystallites.

Finally, we have investigated the possibility of antisite defects in
which a metal atom is substituted for the missing boron octahedron.
In all cases the formation energy is considerably reduced with respect
to that of a B$_6$ vacancy, but still much too high to lead to a measurable
concentration in conditions of thermal equilibrium. What is
remarkable, however, is that the incorporation of either Al or La
completely removes the original, large magnetic moment.
This leads us to the following tentative scenario for the evolution
of the magnetism in La-doped CaB$_6$: As La is incorporated at the
surface of the growing crystal, the resulting local distorsion
increases the probability of formation of B$_6$ vacancies. If the
La concentration in the flux is low, these will remain empty, while
the La atoms substitute for Ca. Provided the voids form connected
clusters as described above, ferromagnetism can occur, with a moment which
increases with the La concentration in the flux. Once the latter
exceeds a certain value, some La atoms will start to ``see'' the B$_6$ vacancies
and occupy them, at a considerable energy gain. At still higher La
concentration in the flux, every created B$_6$ vacancy is occupied by
a La atom, and the magnetization vanishes altogether.

In summary, we have shown that the concentration of point defects
in CaB$_6$ grown under conditions of thermal equilibrium is too small
to account for the transport properties of this material.
Having found that of all intrinsic point defects only the
B$_6$ vacancy carries a (large) magnetic moment, we have postulated
that, due to kinetic effects, such ``holes'' are created during the
growth of the crystals, especially when the lattice is distorted by
large impurities or cation vacancies, and we have presented a
plausible scenario for the
magnetic behaviour of La-doped CaB$_6$ as function of concentration.
The large moment carried by B$_6$ vacancy also leads to a natural
explanation for the high saturation magnetization observed in unetched
polycrystalline samples as a surface effect.

\section*{Acknowledgements}
It is a pleasure to thank A. D. Bianchi for his comments on the manuscript.

\begin{table}
\caption[p]{Formation energy  ($^{a}$from the gas phase, $^{b}$ from 
the elements dissolved in a liquid aluminium flux, $^{c}$ from borothermal
reduction of CaO), and local  magnetic moment of point defects in CaB$_6$}.
\begin{center}
\label{param}
\begin{tabular}{|l|c|c|} 
\hline 
Defect  & E $_{Form}$ [eV] & Moment  [$\mu_B$]  \\
\hline
Ca vacancy & 6.63$^{a}$, 4.96$^{b}$, 4.6$^{c}$ &   $<$0.001 \\
La (Ca)& -1.94$^{a}$, -0.14$^{b}$ & $<$0.001 \\
Al (Ca)& 3.73$^{a}$, 5.0$^{b}$ & $<$0.001 \\
B (Ca)&  4.69$^{a}$, 8.5$^{b}$  & $<$0.001    \\
B vacancy & 11.05$^{a}$, 5.6$^{b,c}$ & $<$0.001 \\
B$_6$ vacancy & 50.7$^{a}$,18.3$^{b,c}$ & 2.36\\
Ca (B$_6$) & 44.2$^{a}$, 13.5$^{b}$ & 1.32 \\
La (B$_6$) & 40.6$^{a}$, 11.7$^{b}$ &  $<$0.001     \\
Al (B$_6$) & 42.5$^{a}$, 13.1$^{b}$ & $<$0.001 \\
\hline
\end{tabular}

\end{center}

\end{table}


\begin{references}
\bibitem{gain} As a comparison, the calculated difference between
the cohesive energies of the two materials is 4.37 ev/formula unit.
\end{references}
\end{document}